\documentclass{article}

\usepackage[english]{babel}

\usepackage[letterpaper,top=2cm,bottom=2cm,left=3cm,right=3cm,marginparwidth=1.75cm]{geometry}

\usepackage{amssymb,amsmath,mathtools,amsthm}
\usepackage{listings}
\usepackage{physics}
\usepackage{hyperref}
\usepackage{multirow}
\usepackage{array} 
\usepackage{tikz} 
\usepackage{authblk}
\usepackage{cite}
\usepackage{booktabs}

\usepackage[labelsep=period]{caption}
\usepackage{subcaption}

\usepackage{array}
\usepackage{tabu}   
\usepackage{xspace} 

\usepackage{graphicx}
\usepackage{float}

\usepackage{booktabs}

\graphicspath{{figures/}}

\usetikzlibrary{automata,arrows.meta,calc,positioning}

\usetikzlibrary{shapes.geometric}
\usetikzlibrary{external}

\title{Classification and transformations of quantum circuit decompositions for permutation operations}
\date{}

\newcommand{\SCNW}{strict clean non-wasting}   
\newcommand{\RCNW}{relative clean non-wasting}  
\newcommand{\SDNW}{strict dirty non-wasting}  
\newcommand{\RDNW}{relative dirty non-wasting}  
\newcommand{\SCWE}{strict clean wasting-entangled}  
  
\newcommand{\SDWE}{strict dirty wasting-entangled}  
  
\newcommand{\SCWS}{strict clean wasting-separable}  
\newcommand{\RCWS}{relative clean wasting-separable}  
\newcommand{\SDWS}{strict dirty wasting-separable}  
\newcommand{\RDWS}{relative dirty wasting-separable}

\newcommand{\C}{\mathbb{C}}
\newcommand{\Hl}{\mathcal{H}}

\author[1]{Ankit Khandelwal$^\dagger$}
\author[2]{Handy Kurniawan$^\dagger$}
\author[3]{Shraddha Aangiras}
\author[4,5,6]{\"Ozlem Salehi}
\author[4,6]{Adam Glos\thanks{aglos@iitis.pl}}
\affil[1]{TCS Research, Tata Consultancy Services, India}
\affil[2]{Facultad de Informática, Universidad Complutense de Madrid, Spain}
\affil[3]{Rashtreeya Vidyalaya Preuniversity College, India}
\affil[4]{Institute of Theoretical and Applied Informatics, Polish Academy of
	Sciences, Poland}
\affil[5]{QWorld Association, Tallinn, Estonia}
\affil[6]{Algorithmiq Ltd, Kanavakatu 3C 00160 Helsinki, Finland}

\begin{document}
\maketitle

\def\thefootnote{$\dagger$}\footnotetext{These authors contributed equally to this work}

\begin{abstract}
Efficient decomposition of permutation unitaries is vital as they frequently appear in quantum computing. In this paper, we identify the key properties that impact the decomposition process of permutation unitaries. Then, we classify these decompositions based on the identified properties, establishing a comprehensive framework for analysis. We demonstrate the applicability of the presented framework through the widely used multi-controlled Toffoli gate, revealing that the existing decompositions in the literature belong to only three out of ten of the identified classes. Motivated by this finding, we propose transformations that can adapt a given decomposition into a member of another class, enabling resource reduction. 
\end{abstract}

\section{Introduction}

The idea of quantum computers was first introduced in Feynman's seminal paper in 1982 \cite{Feynman1982}. Since then, quantum computing has evolved from a theoretical concept to a promising technology with numerous potential applications. The fundamental difference between classical and quantum computing lies in the way information is processed and manipulated. Quantum computers use the principles of quantum mechanics, like quantum superposition, to perform complex calculations faster than classical computers for certain tasks. Over the past few decades, many powerful quantum algorithms, both for fault-tolerant quantum computers ~\cite{grover1996fast,shor1999polynomial} and for noisy intermediate-scale quantum hardware ~\cite{farhi2014quantum, peruzzo2014variational, kadowaki1998quantum}, have been proposed. 

The plethora of developed algorithms is dedicated to classical problems like searching and combinatorial optimization. The best-known algorithm that falls into this class is Grover's Search Algorithm~\cite{grover1996fast}, which relies on the utilization of an oracle to `mark' the desired states within the search space. The oracle implementation usually consists of 3 steps. First, an auxiliary qubit is set to $\ket{1}$ in case the corresponding state represents a searched element, and $\ket{0}$ otherwise. This can be achieved by applying a permutation unitary. Then a $Z$ gate is applied to this auxiliary qubit which results in a phase change if it is set to $\ket{1}$. Finally, the inverse of the permutation is applied. Similar oracle implementations appear not only in search-based algorithms~\cite{grover1996fast, santha2008quantum, magano2022quantum, glos2021quantum, ambainis2019quantum}, but also in Noisy Intermediate-Scale Quantum (NISQ) era combinatorial optimization algorithms like Grover Adaptive Search~\cite{gilliam2021grover}, and variants of Quantum Approximate Optimization Algorithm~\cite{bako2022near, pelofske2023high}. Once again, such algorithms require a permutation to mark the desired states. Additionally, permutations also appear in quantum circuits performing arithmetic operations like in Shor's factorization algorithm~\cite{vedral1996quantum}, in quantum wavelet transform implementation~\cite{fijany1999quantum}, and finally in error mitigation~\cite{botelho2022error}. 

Although such permutations are natural for classical computing, implementing them on a quantum hardware is challenging. The number of qubits involved in a gate in the majority of current gate-based architectures cannot exceed two, which is also anticipated for fault-tolerant quantum computers. Consequently, multi-qubit gates have to be decomposed into ones that are native to the hardware. It is essential that the resulting circuits are not costly in terms of gate count and depth for the following reasons. First of all, the reduction in the number of CNOTs ($T$ gates) for the NISQ (fault-tolerant) era increases the robustness of the quantum systems to noise. Additionally, minimizing the depth of the circuit in the noisy era reduces the effect of the decoherence on the quantum system and decreases the time required for obtaining a single measurement outcome from the circuit.

One may consider various simplifications and assumptions to optimize the decomposition process of permutation gates. For instance, using a `relative-phase' unitary may already lead to a significant reduction in resources, in case the relative phase is not critical~\cite{maslov2016advantages,PhysRevA.52.3457}. Additionally, when applying permutation gates right before the phase change in the oracular algorithms, one can easily remove the gates that are not acting on the qubit on which the phase-changing gate is applied. This simplification will result in a different permutation, but eventually, the oracle will produce the same quantum state provided that uncomputation is performed symmetrically. In the case of auxiliary qubits, if these simplifications leave the auxiliary qubits in a state different than $\ket{0}$, one can still consider a `dirty' implementation for the consequent permutations, in which the auxiliary system can be in an arbitrary state. Furthermore, particularly cumbersome (usually defined over many qubits) unitaries can be implemented using less number of gates at the cost of potentially `wasted' auxiliary qubits, i.e. the implementation does not act as identity on the auxiliary system. These simplifications combined with others, expand the possibilities of transpiling the circuits, ultimately enhancing the effectiveness of the decomposition process. Similar ideas are investigated in~\cite{amy2021phase}, where the authors explore relative phase circuit constructions for classical functions. The authors also take into account the state of the auxiliary system and provide efficient constructions of some classical functions like the multi-controlled Toffoli (MCT) gate.

In this paper, we make the following contribution. First, we identify the various properties of the permutation unitaries that critically impact their decomposition into quantum circuits, taking into account the aforementioned simplifications and assumptions. Then, based on the identified properties, we classify the unitaries and thus the corresponding quantum circuits, providing a formal mathematical definition for each class. Despite naturally 12 classes being introduced for the classification of the generalized unitaries, we find out that only 10 of them are unique. To demonstrate this classification, we examine the existing implementations of the multi-controlled Toffoli gate and determine the class each implementation belongs to. Our review shows that, even for this widely considered significant operation, implementations exist only for three of the introduced classes. Finally, we propose several transformations, that enable the conversion of a member of one class to a larger class within the classification, enabling reduction of the resources required. Our contribution not only provides a comprehensive overview and standardization of unitary permutation decompositions but also serves as a valuable resource for those new to designing and preparing quantum circuits offering a solid introduction to the topic and facilitating efficient circuit design.

The paper is organized as follows. In Sec.~\ref{sec:classification},
 we introduce the concept of permutation and generalized permutation matrices, and their classification depending on their particular properties. In the same section, we present a review of the implementations of multi-controlled Toffoli. In Sec.~\ref{sec:transformations},
we introduce the class-preserving transformations for quantum circuits. Finally, in Sec.~\ref{sec:conclusions}, we conclude and discuss our results together with their potential applications.

\section{Classification of permutation unitaries}\label{sec:classification}

In this section, we identify the properties significantly impacting the decomposition of unitaries and present the unique classes based on those properties. In addition, we classify existing multi-controlled Toffoli implementations into the proposed classes.

\subsection{Preliminaries}

A unitary matrix $U\in {\mathrm U}(\C^N)$ is called a \textit{permutation unitary} if there exists a permutation $\pi: [N] \to [N]$ s.t. $\bra {\pi(j)} U \ket j =1$, where $[N] \coloneqq \{0,\dots,N-1\}$. Permutation unitaries have the property that when they act on the computational basis state $\ket{i}$, they transform it to a computational basis state $\ket{j}$. Note that in this paper we consider only algebraic transformations for simplicity, so removing a global phase is not allowed unless it is explicitly stated. On the other hand, \emph{generalized permutation unitaries} satisfy $|\bra {\pi(j)} U \ket j| =1$ and they transform a basis state into another basis state up to a phase. It is efficient to simulate the evolution composed of such unitaries acting on a basis state $\ket{i}$, as it is equivalent to applying the corresponding permutation on $i$. Hence, there is a direct correspondence between such a quantum evolution and classical reversible computing. 

Depending on the particular unitary, one may consider various quantum circuits. For example, to implement a Toffoli gate using only CNOTs and 1-qubit gates, one can choose the well-known implementation consisting of six CNOTs~\cite{nielsen2010quantum}, which is known to be tight~\cite{shende2008cnot}. Alternatively, one can use Sleator-Weinfurter construction that uses five 2-qubit gates \cite{sleator1995realizable}. When implementing such permutations, one may consider different aspects of the unitary. First and foremost, allowing a change in phase may result in a significant reduction in resources. We will call such implementation a \emph{relative-phase implementation} where the corresponding unitary will be a generalized permutation matrix. For example, while the Toffoli gate requires exactly 6 CNOTs to be implemented, its relative-phase version requires only 3 CNOTs~\cite{divincenzo1994results}. Hence we can deduce that by relaxing the condition of the matrix being 0-1, we can find a less costly circuit. Using such relative phase implementations turns out to be particularly useful in various applications, which include optimization of implementation of multi-controlled Toffoli gate~\cite{maslov2016advantages}, uncomputation~\cite{amy2021phase}, efficient implementation of quantum finite automata~\cite{birkan2021implementing} and Grover's search algorithm~\cite{zhang2022quantum}.

Relaxing the condition into general local phases is not the only aspect that can be considered. In particular, implementation of the unitaries often requires an auxiliary system consisting of qubits that do not store input, nor the output of the permutation. In case it is assumed that the auxiliary system should be initialized to state $\ket{0}$, the implementation is considered \emph{clean}. However, such auxiliary qubits may not always be available. For instance, one may want to implement a permutation in the middle of the quantum circuit, on a quantum device with sparse connectivity: in such a case, passing clean qubits may require extra SWAPs, which are costly operations. Therefore, it becomes desirable to have an implementation that is robust even when auxiliary qubits are not in an ideal state, which we refer to as a \emph{dirty} implementation. However, by relaxing the condition imposed on the input state of the auxiliary system, one should expect that quantum circuits may require more quantum resources.

The final aspect concerns the output state of the auxiliary system. In the \emph{wasting} setting, the auxiliary qubits are left in a different state than given initially, while in the usual \emph{non-wasting} setting, we expect the output state to be identical to the input state. Formally, the non-wasting setting is defined as acting with an identity on the auxiliary system.  While we call such an approach wasting, note that those qubits can be reused as dirty auxiliary qubits in the upcoming part of the quantum circuit. In this case, one may consider two kinds of wastes: provided the auxiliary and main systems were initially uncorrelated, the wasted auxiliary system might be left uncorrelated to the main system, or it may be entangled with it in general. We will refer to those cases as \textit{wasting-seperable} and \textit{wasting-entangled} respectively.

\subsection{Classification and verification of generalised permutations} \label{sec:classes}

Let us consider a permutation $\pi$ and the corresponding 0\,-1 permutation unitary $U_\pi$. In this subsection, we will investigate and classify unitary operations corresponding to different implementations of $U_\pi$, taking into account various aspects we discussed in the previous subsection. Formally, we will consider classes of unitaries $\mathcal{C}_\beta = \{U \in  {\mathrm U}(\Hl_M \otimes \Hl): \beta(U)\}$, where $\Hl_M$ is the system on which $\pi$ is acting, $\Hl$ is the auxiliary system, and $\beta$ denotes extra conditions to be satisfied by $U$. Note that $\Hl$ is an arbitrary size register -- hence we can think of the class $\mathcal{C}_\beta$ as `all unitaries implementing $\pi$ with a given auxiliary system, satisfying condition $\beta$'. As an extreme and special case, the auxiliary system can be 1-dimensional. 

The classes presented here are formally defined for unitary matrices. However, since each quantum circuit is in fact a realization of some particular unitary, the classification naturally extends to the quantum circuits as well. Besides classification, we will also present methods to verify if a given unitary belongs to a particular class.

The first class is the \emph{\SDNW}. In this class, we allow arbitrary input state $\ket{\psi}$ for the auxiliary, but we require the output state to be the same. Therefore such $U$ should satisfy
\begin{equation}
    U \ket{b}\ket{\psi} = \exp(i \varphi) \ket{\pi(b)} \ket{\psi}
\end{equation}
for any $\ket{b}$, for some fixed real $\varphi$. Note that throughout the paper we always mean algebraic equality, not up to the global phase. It is straightforward to see that $U$ is a product matrix s.t. $U = \exp(i \varphi) U_{\pi} \otimes I$. To verify if a given matrix $U$ belongs to this class, it is enough to check whether $\langle U(U_\pi ^\dagger \otimes I)\rangle  = I$. where $\langle A\rangle \coloneqq \overline {A_{00}} \cdot A$.

The natural extension is the \emph{\RDNW} class in which we are acting with some generalized permutation unitary on $\Hl_M$. In such a case, the unitary satisfies
\begin{equation}
    U \ket{b}\ket{\psi} = \exp(i \varphi_b)\ket{\pi(b)} \ket{\psi}
\end{equation}
for some $\varphi_b$ depending on the computational basis state $\ket{b}$. In such a case, we can write $U = \tilde U_{\pi} \otimes I$, where $\tilde U_{\pi}$ is a generalized permutation matrix. This observation gives a natural verification procedure to check if a given $U$ belongs to the class: it is enough to apply inverse Kronecker product to decompose $U$ into a product of two unitary operations $V\otimes W$ with dimension of $V$ being the same as for $U_\pi$. Note that if the procedure fails, then it is immediately clear that $U$ cannot belong to this class. On the other hand, if such a decomposition exists, then it is unique up to the global phase acting on both $V$ and $W$, see Appendix~\ref{sec:reverse-kronecker-product} for a detailed explanation. If the procedure succeeds in finding decomposition $U=V\otimes W$, it is enough to verify $ |V| U_\pi^\dagger  = I$ and $\langle W \rangle = I$.

Note that the cases in which $\Hl$ is one-dimensional naturally fall into the categories above, because then taking tensor product with $1\times 1$ identity matrix does not change the unitary.

The next class is \emph{\SCNW}, where we assume that the auxiliary system is in state $\ket{0}$. Hence, any unitary $U$ belonging to this class should satisfy
\begin{equation}
    U \ket{b}\ket{0} = \exp(i \varphi) \ket{\pi(b)} \ket{0}.
\end{equation}
for any $\ket{b}$ and some real value $\varphi$. From the condition above, it is clear that $(I \otimes \bra{0}) U  (I \otimes \ket{0}) = \exp(i \varphi) U_\pi$. In order to verify whether a unitary $U$ belongs to this class, we need to first identify the corresponding entries of $U$ for the auxiliary system being in $\ket{0}$:
\begin{equation}\label{eq: uprime}
    U' \coloneqq ( I\otimes \bra{0}) U (  I\otimes \ket{0}).
\end{equation}
Then, we can check whether $\langle U' U_\pi^\dagger \rangle$ is the identity matrix. For the relative phase version, \emph{\RCNW}, it is enough to replace $U$ with $|U|$ in Eq.~\eqref{eq: uprime}. Note that we cannot tell anything about how these unitaries act on a quantum state which is not of the form $\ket{\psi}\ket{0}$.

While the {\SDNW}  can be viewed as a generalization of the clean auxiliary system in the sense of the input state, we can also consider a generalization in the context of the output state. Let us continue with the class \emph{\SCWE}. Unitaries belonging to this class may change $\ket 0$ auxiliary state into some arbitrary quantum state, i.e.
\begin{equation}
    U \ket{b}\ket{0} = \ket{\pi(b)} \ket{\psi_b}. \label{eq:strict-clean-wasting-entangled-state-action}
\end{equation}
Since we allow entanglement between the systems, $\ket{\psi_b}$ may depend on $b$. Note that strict and relative-phase versions correspond to the same class, as the local phase can be hidden inside the state of the auxiliary system:
\begin{equation}
    U \ket{b}\ket{0} =   \exp(i \varphi_b) \ket{\pi(b)}\ket{\psi_b} =  \ket{\pi(b)} ( \exp(i \varphi_b) \ket{\psi_b}) =  \ket{\pi(b)}\ket{\psi'_b}.
\end{equation}
Note that for a similar reason, we do not need to write a global phase explicitly in Eq.~\eqref{eq:strict-clean-wasting-entangled-state-action} contrary to the previously introduced strict classes.

Now, let us discuss how one can verify if a given $U$ belongs to this class. Note that since $\ket{\psi_b}$ should be a quantum state for any $b$, it is necessary that it is normalized. Equivalently, we can express $\ket{\psi_b}$ as
\begin{equation} \label{eq: cew}   
\ket{\psi_b} \coloneqq  (\bra{b} \otimes  I)(U_\pi^\dagger\otimes I) U ( \ket{b} \otimes\ket{0}) = (\bra{\pi(b)} \otimes  I) U ( \ket{b} \otimes\ket{0})
\end{equation}
and it is necessary to have 
\begin{equation}
    \| \ket{\psi_b} \|_2 =  \braket {\psi_b} = 1
\end{equation}
for all $\ket b$. In fact, this condition is also sufficient to test if a given $U$ belongs to this class. Let us suppose that $V$ is an arbitrary unitary operator. $V$ can always be written as 
\begin{equation}
    V = \sum_{b} \sum_c \ketbra{\xi_{b,c}}{b,c}
\end{equation}
where $ \ket{\xi_{b,c}} $ is the state obtained by applying $V$ on $\ket{b,c}$.
	Note that using Eq.~\eqref{eq: cew}
	\begin{equation}
	1 = \braket{\psi_b}{\psi_b} = \bra{\psi_b} (\bra{b}U_{\pi}^\dagger\otimes  I)  V (\ket{b} \otimes\ket{0}).
	\end{equation}
	Hence
	\begin{equation}
	\ket{\xi_{b,0}}= V (\ket{b} \otimes\ket{0}) = (U_{\pi}\otimes I)(\ket{b} \otimes \ket{\psi_b}).
	\end{equation}
	So we have 
	\begin{equation}
	    V = \sum_{b} \sum_{c\neq 0} \ketbra{\xi_{b,c}}{b,c} + \sum_{b}  (U_{\pi}\otimes I)\ketbra{b,\psi_b}{b,0},
	\end{equation}
	and finally 
	\begin{align}
	    V \ket{b}\ket{0} &=  (U_{\pi}\otimes I)\ket{b,\psi_b} \braket{b,0}{b,0}= (U_{\pi} \ket{b}) \otimes \ket{\psi_b} =  \ket{\pi(b)} \otimes \ket{\psi_b},
	\end{align} 
	which shows that the necessary requirement is in fact also sufficient.

In case there is no entanglement between the input and auxiliary systems in the {\SCWE} case, we obtain \emph{\SCWS} class. Suppose that there is a fixed $\ket{\psi}$ s.t. for any $\ket{b}$ we have
\begin{equation}
    U \ket{b}\ket{0} = \ket{\pi(b)} \ket{\psi}.
\end{equation}
We can design a similar test to decide if a given $U$ belongs to the class, as we had for {\SCNW}; however, here we have to make sure the output state is the same for each input state. 
Let us define
\begin{equation}
\ket{\phi_b} = 	(\bra{b} \otimes I)(U_{\pi}^\dagger\otimes I) U ( \ket{b} \otimes\ket{0}).
\end{equation}
Then, it is enough to check that for any $\ket b$
\begin{equation}
    \braket{\phi_b}{\phi_0} = 1,
\end{equation}
which is the necessary and sufficient condition to test whether a given $U$ belongs to this class. Note that in general {\SCWS} unitary might not be separable itself -- this is because we make no assumption on how the unitary acts on any other basic state than $\ket{0}$ of the auxiliary system, which itself may introduce entanglement. Finally, since there is no entanglement between main and auxiliary subsystems, one can easily recover the states of auxiliary qubits by resetting them at the end. 

For the wasting-separable condition, we cannot move the relative phase to the auxiliary system as it was done for the entangled case. Hence relaxing the class to the relative phase one results in \emph{\RCWS}. Any unitary $U$ from this class satisfies
\begin{equation}
    U \ket{b}\ket{0} = \exp(i \alpha_b) \ket{\pi(b)} \ket{\psi}
\end{equation} 
for some real $\alpha_b$. Similarly, as for the clean case, we can recover states of auxiliary qubits by resetting them at the end. Since $\ket{\psi}$ is the same for each input state, one can use the following to test whether $U$ belongs to the discussed class:
\begin{equation}
    |(I\otimes \bra{\psi}) U (I \otimes \ket {0})| = U_\pi. 
\end{equation}

Now, let us move to dirty wasting classes and start with \emph{\SDWE} class. The state of the auxiliary system does not need to be preserved, and in fact, the new state may depend on the global input state of the main and auxiliary systems. Formally, for a general quantum state $\ket{\phi}$, this can be expressed as
\begin{equation}
    U \ket{b,\phi} = \ket{\pi(b), \psi_{b, \phi}}.
\end{equation}
By linearity of unitary $U$, it is sufficient for the expression to hold only for all basis states $\ket{c}$:
\begin{equation}\label{eq: DEWS}
    U \ket{b,c} = \ket{\pi(b), \psi_{b, c}}.
\end{equation}
  Any unitary $U$ belonging to {\SDWE} class can be seen as `clean' wasting-entangled for any basis input state $\ket{c}$ of the auxiliary system. Therefore, one could directly extend Eq.~\eqref{eq: cew} by expressing $\ket{\psi_{b,c}}$ as 
\begin{equation}    
\ket{\psi_{b,c}} \coloneqq  (\bra{b} \otimes  I)(U_\pi^\dagger\otimes I) U  \ket{b,c},
\end{equation}
and the following should be true:
\begin{equation}
    \| \ket{\psi_{b,c}} \|_2 =  \braket {\psi_{b,c}} = 1
\end{equation}
for all $b,c$. Similarly, as it was for {\SCWE}, one can show there is no notion of a strict or relative phase.

Finally, let us consider the \emph{\SDWS} class. Similarly to the entangled case, any such $U$ should satisfy
\begin{equation}
    U \ket{b,\phi} = \ket{\pi(b), \psi_\phi}.
\end{equation}
Note that the dependency on $b$ for $\psi$ is dropped. One can immediately see that, since there is no correlation between the states, $U$ must be a separable unitary, i.e. $U = V \otimes W$ for some matrix $W$.  Once we get this decomposition using e.g. the method presented in Appendix~\ref{sec:reverse-kronecker-product}, it is enough to check whether 
\begin{equation}
 \langle V U_\pi^\dagger  \rangle= I.   
\end{equation}
Similarly for the \emph{\RDWS} class, one needs to check if $|V| U_\pi^\dagger = I$.

\begin{figure}[t]
\centering
\includegraphics[scale=0.9]{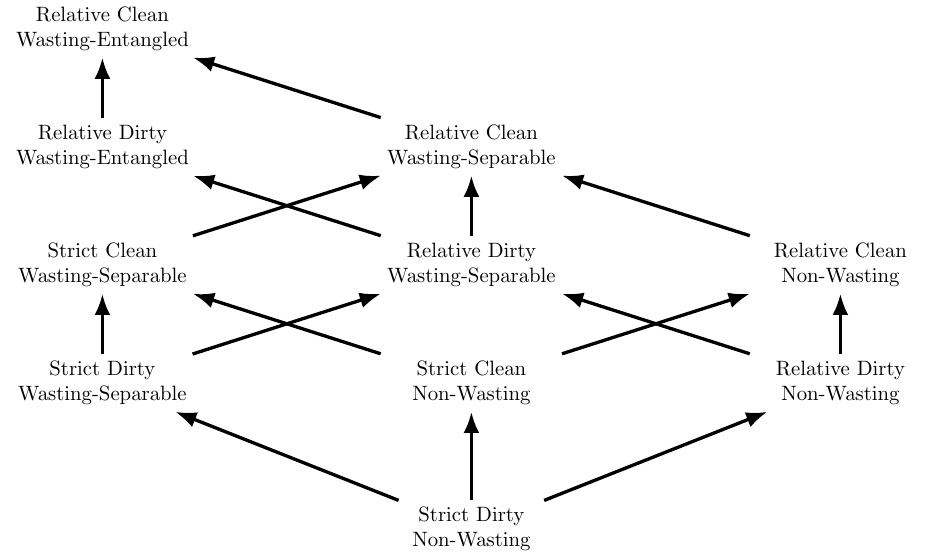}
\caption{Partial order of the classes introduced in the Sec.~\ref{sec:classes}.}

\label{fig:dag}
\end{figure}

We summarize all the classes introduced so far in Table.~\ref{tab:class_verification}, and we present the relationship between them in Fig.~\ref{fig:dag}. The arrows indicate inclusion. Consider a unitary corresponding to a dirty implementation. Such unitary will also work with a clean auxiliary but the converse is not true. Hence, we can conclude that classes with clean auxiliary systems are supersets of the corresponding dirty classes. A similar relationship exists between the strict and relative-phase classes, as any strict implementation can be considered a relative-phase implementation as well, but the converse is not true. Finally, a unitary corresponding to a non-wasting implementation also belongs to the corresponding wasting class. Therefore, we have an inclusion relation between the non-wasting and wasting classes.

\begin{table}[h]
\setlength\extrarowheight{4pt}
\centering
\begin{tabular}{@{}lm{5.5cm}m{4.5cm}@{}}
\toprule
\textbf{Class} & \multicolumn{1}{l}{\textbf{Supplementary definitions}} & \multicolumn{1}{l}{\textbf{Verification}} \\ 
\addlinespace[0.2em]
\hline 
\midrule
S-C-NW &     $U' \coloneqq ( I \otimes \bra{0}) U (  I\otimes \ket{0})$       & $\langle U' U_\pi^\dagger \rangle =  I$ \\
\midrule %
R-C-NW &     $U' \coloneqq ( I \otimes \bra{0}) U (  I\otimes \ket{0})$       & $|U'| U_\pi^\dagger  =  I$ \\ \midrule 
S-D-NW &    ---        & $\langle U (U_\pi^\dagger \otimes I) \rangle =  I$ \\ \midrule
R-D-NW          &    $U= V\otimes W$  & $  |V| U_\pi^\dagger  = I \land \langle W \rangle = I$\\ \midrule
S-C-WE           &  
\multirow{2}{*}{$\ket{\psi_b} \coloneqq (\bra{\pi(b)} \otimes  I) U 
( \ket{b} \otimes\ket{0}) $} &
\multirow{2}{*}{$\forall b\   \braket{\psi_b} = 1$  } \\ 
R-C-WE          &  & \\ \midrule
S-D-WE         &  
\multirow{2}{*}{$\ket{\psi_{b,c}} \coloneqq  (\bra{\pi(b)} \otimes  I) U  \ket{b,c}$} &
\multirow{2}{*}{$\forall b,c\   \braket{\psi_{b,c}} = 1$   } \\ 
R-D-WE          &  & \\ \midrule

S-C-WS &     $\ket{\phi_{b}} \coloneqq  (\bra{\pi(b)} \otimes  I) U ( \ket{b} \otimes\ket{0})  $       &    $\forall b   \braket{\phi_{b}}{\phi_{0}}=1 $        \\ \midrule 
R-C-WS          &    
$\ket{\psi} \coloneqq (\bra{\pi(0)} \otimes  I) U ( \ket{0} \otimes\ket{0})$       & $|(I\otimes \bra \psi) U ( I \otimes \ket 0)| = U_\pi$  \\ \midrule
S-D-WS        &     $U= V\otimes W$ & $ \langle V U_\pi^\dagger \rangle = I$ \\ \midrule
R-D-WS         &     $U= V\otimes W$       & $|V| U_\pi^\dagger = I$ \\ 
\bottomrule

\end{tabular}
\caption{The list of all classses introduced in Sec.~\ref{sec:classes} and their verifications. Here we assume $|A|$ to be the element-wise absolute value of elements, and $\langle A\rangle\coloneqq A \cdot \overline{A_{00}}$. The purpose of the first is to remove relative phases in the unitary, while the latter is to remove the global phase. In all cases, $U$ is the matrix being tested, and $U_\pi$ is the true 0-1 unitary matrix of the permutation. Symbols $b$ ($c$) denote an element of the computational basis for the main (auxiliary) system. The decomposition $U=V\otimes W$ with unitaries $V$, $W$ can be found using the method presented in Appendix~\ref{sec:reverse-kronecker-product} if and only if it exists. If such decomposition does not exist, it means the unitary does not belong to the corresponding class. Note that if $U$ is unitary, one can always find $V,W$ such that they are also unitary. We used the following acronyms for classes: `S' for strict, `R' for relative-phase, `C' for clean, `D' for dirty, `NW' for non-wasting, `WS' for wasting-separable, and `WE' for wasting-entangled.}
\label{tab:class_verification}
\end{table}

\subsection{Example -- MCT classification} \label{sec:mct-review}

In this part of the section, through an example of multi-controlled Toffoli operation, we show exemplary quantum circuits belonging to the classes defined above. We start with a brief review of the most popular and interesting from the perspective of this paper decompositions of this gate, and we end up attaching appropriate minimal classes to them in the sense of the order defined in Fig.~\ref{fig:dag}. As part of the conclusions, we notice that there are many classes for which there is no implementation in the found literature. 

The seminal work on this topic, Barenco et al. \cite{PhysRevA.52.3457}, gives multiple decompositions for the MCT gate that require $0,1$ or $n-2$ auxiliary qubits where $n-1$ is the number of control qubits. These decompositions form the basis of various other works and are available for use in widely used quantum computing libraries like Qiskit \cite{Qiskit}. While the final implementations are strict, the building blocks are relative phase Toffoli gates with a CNOT count equal to 3.
While in~\cite{PhysRevA.52.3457} the decompositions are given for a particular number of auxiliary qubits, Baker et al.
\cite{baker2019decomposing} provide decompositions for an arbitrary number of auxiliary qubits, for both clean and dirty auxiliary qubits cases.
In \cite{10.1145/1366110.1366168}, the authors look at pairs of gates that result in fewer elementary gates when decomposed together.
In \cite{linearDepthSilva}, Silva and Park present a linear depth decomposition of $C^nU$ which uses no auxiliary qubits. They claim lower depth for $n>6$ compared to other methods.
Balauca and Arusoaie \cite{10.1007/978-3-031-08760-8_16}, provide decompositions with logarithmic depth and a logarithmic number of auxiliary qubits. They also suggest an improvement to one of the currently available methods in Qiskit. They provide constructions for relative-phase Toffoli controlled by 2 or 3 qubits and use these gates as building blocks for their MCT decompositions.
In \cite{maslov2016advantages}, Maslov gives examples of replacing suitable pairs of MCT gates with their relative phase implementations. For example, an $n-1$ controlled MCT gate $(n \geq 5$) uses $\lceil \frac{n-3}{2} \rceil$ auxiliary qubits and $8n-16$ $T$ gates, $28n-20$ CNOT gates, and $4n-10$ Hadamard gates. The implementations in this paper are the ones available in Qiskit under the names \texttt{RCCXGate} (relative phase Toffoli with 2 controls), \texttt{RC3XGate} (relative phase Toffoli with 3 controls), and \texttt{MCXVChain} (strict MCT with both clean and dirty auxiliaries).

In the literature, there exist also studies focusing on providing decompositions oriented towards fault-tolerant computing. Biswal et al. \cite{PhysRevA.100.062326} uses the Clifford + $Z_N$ gate library to give a decomposition with linear phase depth and quadratic phase count while requiring no auxiliary qubit. Here $Z_N\coloneqq\sqrt[N]{Z}, N=2^n$. The authors also give a unit phase depth decomposition with exponential phase count and auxiliary qubits. Jones \cite{PhysRevA.87.022328} presents a fault-tolerant Toffoli gate that can detect a single $Z$ error occurring with some fixed probability in any one of the eight $T$ gates required to implement the Toffoli gate and improves upon the $T$ gate count required for the decomposition of MCT by giving a decomposition with four $T$ gates. Finally, Selinger \cite{PhysRevA.87.042302}  gives a decomposition of the Toffoli gate with $T$ depth one, and shows that $m$ additional controls can be added to it at the cost of linear $T$ count, and logarithmic $T$ depth. 

In the papers considered till now, some authors have focused on decomposing MCT into Toffoli or relative phase Toffoli gates, while others used Clifford + $T$ or even  Clifford + $Z_N$. The following papers go in a different direction, using the NCV (`N' from NOT, `C' from CNOT, and `V' from $\sqrt X$) and its derivative gate libraries. Note here that many of the IBM quantum devices have NCV gates within their basis gate set.
In \cite{5291355} and \cite{5954249}, the NCV gate set is used to reduce the gate count for decomposing MCT with an arbitrary number of auxiliary qubits.
Kole and Datta \cite{7884793} provide some improvements in terms of gate counts. From an NCV decomposition, one can go to an NCVW (W being $\sqrt[4]X$) decomposition to reduce the number of gates even further~\cite{7435022}.
Maslov et al. \cite{4378213} provide various useful simplifications for the decomposition of MCT gate to lower the gate count.
In \cite{6721034}, the authors define and use generalized Peres gates to construct MCT decompositions without the need for auxiliary qubits which have a lower gate count in the NCR$_k$ ($R_k\coloneqq\sqrt[k]{X}$) gate set. The gate count can be further reduced in the presence of auxiliary qubits. In \cite{7424064}, the authors show how adjacent Toffoli-CNOT or Toffoli-Toffoli gates can be decomposed together using NCV to minimize gate count. Finally, a similar concept is used in \cite{Szyprowski2012} but with pairs of larger gates that are decomposed together to reduce the gate count using the NCVW gate library.

In Table~\ref{tab:mct_summary}, we present the list of decompositions of MCT mentioned above together with the minimal class each decomposition belongs to. As we can see, only three classes have their representatives, namely: {\SCNW}, {\SDNW}, and {\SCWE}. 

\begin{table}[]
\setlength\extrarowheight{2pt}

\centering
\begin{tabular}{@{}llp{8cm}@{}}
\toprule
\textbf{Class}                & \textbf{Implementations}                               & \textbf{Notes}                                               \\
\addlinespace[0.2em]
\midrule
\multirow{11}{*}{S-D-NW} & Barenco et al. \cite{PhysRevA.52.3457} Lemmas 7.2 \& 7.3 & Relative-phase Toffoli gates are used as building blocks for optimal cost. \\
& Barenco et al. \cite{PhysRevA.52.3457} Lemma 7.5                 &  Zero auxiliary qubit construction. \\
& Silva and Park \cite{linearDepthSilva}                 &  Linear depth circuit with no auxiliary. \\
& Maslov \cite{maslov2016advantages}                                       & Implemented in Qiskit as \texttt{MCXVChain} with \texttt{dirty\_ancillas} set to \texttt{True}. Uses relative-phase Toffoli gates as building blocks.\\ 
& Baker et al. \cite{baker2019decomposing} Sec. 2.2                 &  Provides both multi-control and multi-target constructions. \\
& Balauca et al. \cite{10.1007/978-3-031-08760-8_16} Sec. 4.1               & Single auxiliary qubit construction. \\
& Miller \cite{5291355}                                        & Uses NCV gate library for 1 to $n-2$ auxiliary qubits for an MCT controlled with $n$ qubits.  \\
& Miller et al. \cite{5954249} & Uses NCV gate library. The number of auxiliary qubits is optimized and using more does not decrease the depth.\\
& Kole and Datta \cite{7884793}                & Uses NCV gate library. The number of auxiliary qubits is optimized and using more does not decrease the depth.\\
& Biswal et al. \cite{7435022}& Uses NCVW gate library.\\
& Szyprowski and Kerntopf \cite{6721034}       & Uses $\text{NCR}_k$ gate library. \\ 
\addlinespace[0.4em]
\midrule
R-D-NW             & Maslov \cite{maslov2016advantages}   Figures 3 and 4                   & Implemented in Qiskit as \texttt{RCCXGate} and \texttt{RC3XGate}. These are relative phase Toffoli gates with no auxiliary qubits.\\
\addlinespace[0.4em]
\midrule
\multirow{7}{*}{S-C-NW} & Barenco et al. \cite{PhysRevA.52.3457} Lemma 7.11         & Single auxiliary qubit construction. Relative-phase Toffoli gates are used as building blocks for optimal cost.\\
& Maslov \cite{maslov2016advantages}                                                 & Implemented in Qiskit as \texttt{MCXVChain} with \texttt{dirty\_ancillas} set to \texttt{False}. Uses relative-phase Toffoli gates as building blocks.\\
& Biswal et al. \cite{PhysRevA.100.062326} & Fault tolerant using Clifford + $Z_N$ gate library. Unit phase depth circuit. \\
& Baker et al. \cite{baker2019decomposing} Sec. 2.1 &  Provides both multi-control and multi-target constructions. \\
& Balauca et al. \cite{10.1007/978-3-031-08760-8_16}                & Log-depth construction. Uses relative-phase Toffoli gates as building blocks. \\
& Jones \cite{PhysRevA.87.022328} & Fault-tolerant decomposition.\\
& Selinger \cite{PhysRevA.87.042302} & Unit $T$-depth construction. \\ 
\addlinespace[0.4em]
\midrule
S-C-WE             & Selinger \cite{PhysRevA.87.042302}                      & Authors mention that removing the uncomputation of auxiliary qubits can reduce the gate count. \\
\addlinespace[0.4em]
\bottomrule
\end{tabular}%
\caption{The list of decompositions of MCT mentioned in Sec.~\ref{sec:mct-review} together with the minimal class each decomposition belongs to.}
\label{tab:mct_summary}

\end{table}

\section{Class-preserving transformations for quantum circuits}\label{sec:transformations}

As mentioned before, the quantum circuits representing a particular permutation can be classified analogously based on the classes of the corresponding unitaries. We usually prefer to stick to a class because of some restrictions or freedom we encounter, such as whether the auxiliary qubits are clean or dirty, whether they can be wasted, and whether the relative phase matters for the computation. However, once the required class is selected, our next goal is to choose an efficient implementation of the unitary with the goal of minimizing the number of gates, auxiliary qubits, or depth. Therefore, given a particular implementation, we can ask whether there are steps that can \emph{transform} it in a way that the class requirements will still be met, but will save some of the mentioned resources. For a set of quantum circuits $A$ representing a particular class, we define transformation $T$ as a class-preserving map $T:A \to A$. In practice, it is more convenient to define transformations as sets of rules describing how to modify the circuit (if applicable).

Despite the transformations being formally defined to be class-preserving, their particular application becomes apparent when they are applied to a strict subclass. Suppose that we apply transformation $T:A\to A$ on the quantum circuit $q \in B \subset A$, where $B$ is a set of quantum circuits representing a strictly smaller class in the sense of the order presented in Fig.~\ref{fig:dag}. There is no guarantee that $T(q)\in B$, however, one can guarantee by construction that $T(q) \in A$. In other words, if we have a quantum circuit that belongs to the smaller class, after applying such a transformation we can only guarantee that the output will be a member of the larger class $A$. Note that for a permutation it might be simpler to find a quantum circuit with less resources in a larger class. 
 
The same transformations can be also used for a larger class. For example, one can extend this into a larger class $C \supseteq A$ by setting $T'(q) = q$ for any $q\in C\setminus A$ and $T'(q)= T(q)$ otherwise. However, since we are going to define $T$ through a set of simple rules, one would expect a different $T''(q):C\to C$ which acts the same as $T$ on elements of $A$, but provides some meaningful transformation for $q \in C \setminus A$ by applying the same rules on both. In this case, we cannot even guarantee that $T''$ will be a $C$-preserving transformation, as in general, this may depend on the form of the transformation and $C$ itself.

\begin{figure}
    \centering
    \includegraphics[height = 5.5cm]{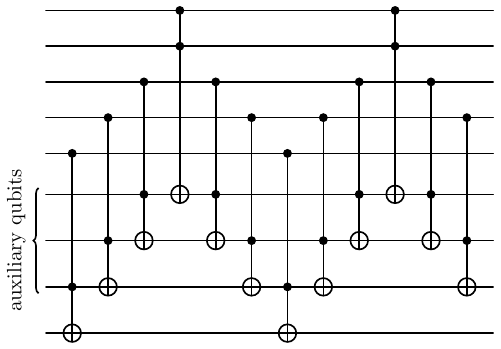}
    \caption{Decomposition of MCT with 5 controls described in Lemma~7.2 from~\cite{PhysRevA.52.3457}.}
    \label{fig:barenco-original}
\end{figure}

We would like to mention, that the idea of transformations was already present in the literature~\cite{amy2021phase}, although they were not defined formally. In this section, we would like to recall some of the existing transformations and introduce new ones that allow for the reduction of resources. We will demonstrate these on the MCT decomposition described in Lemma~7.2 from~\cite{PhysRevA.52.3457}, which belongs to the class {\SDNW}. The decomposition for MCT with 5 controls is presented in Fig.~\ref{fig:barenco-original}. Since this is the smallest class in inclusion, this pedagogical example will effectively convey how our transformation works.

The simplest transformations are those that transform a circuit implementing a strict unitary into a relative-phase one. Consider a subcircuit appearing at the beginning or at the end of a circuit, implementing permutation unitary $U$. One can replace $U$ with its relative-phase version $\tilde{U}$ given that the subcircuit corresponding to $\tilde{U}$ is more resource efficient. Furthermore, if the original circuit contains consecutive subcircuits corresponding to permutations, all such subcircuits can be replaced with their relative-phase versions, whether starting from the beginning or the end. The resulting circuit will be the relative-phase version of the original one. As an example, for the circuit presented in Fig.~\ref{fig:barenco-original} each Toffoli gate can be replaced with its relative-phase variant, saving 3 CNOTs per gate using implementation from~\cite{PhysRevA.52.3457}. Similar replacements of strict permutation gates with relative-phase were already proposed in~\cite{PhysRevA.52.3457,maslov2016advantages,10.1007/978-3-031-08760-8_16}, however in there authors eventually obtained a strict unitary again. Note that since the phase introduced by $\tilde{U}$ may depend on the state of both the auxiliary and the main systems, this might result in additional correlation between them, which makes the transformation preserving for relative clean non-wasting class and all wasting entangled classes. The reason why it is relative clean non-wasting results from the verification procedure presented in Tab.~\ref{tab:class_verification}. Note that replacing strict permutations with relative ones is equivalent to changing 0-1 matrix $U$ into $DU$ or ($UD$) for some diagonal unitary $D$. Therefore, $(I \otimes \bra 0 ) DU (I \otimes \ket 0) = D' U' $ for some different diagonal unitary $D'$, and eventually $|D'U'| U_\pi ^\dagger = |U'|U_\pi ^\dagger$.

Before we continue, we need to introduce the concept of controlled unitaries. Suppose we have a two-system unitary acting on $\Hl_B \otimes \Hl_C$. Then we call unitary $U$ a $B$-controlled unitary if it is of the form
\begin{equation}
    U = \sum_{b=0}^{\dim(B)-1} \ketbra{b} \otimes U_b \label{eq:controlled-def}
\end{equation}
for some unitary operations $U_b$. Note that we can impose additional constraints on $U_b$, in particular, if $U_b$ are (relative-phase) permutations, $U$ may belong to one of the classes introduced in Sec.~\ref{sec:classes}.

\begin{figure}[t]
   \centering

     \includegraphics[height = 5.5cm]{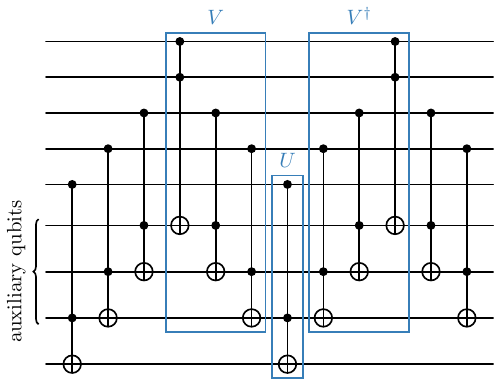}
   \caption{
   Subcircuits corresponding to scenario $VUV^\dagger$ are highlighted. The qubits on which $V$ acts either serve as control qubits for $U$ or are disregarded when applying $U$. The subcircuits corresponding to unitary $V$ and $V^\dagger$ can be replaced with their relative-phase versions to result in a dirty non-wasting preserving transformation.
}
   \label{fig:vuvdagger}%
\end{figure}

We continue with a type of transformation that is class-preserving even for the smallest class which is the {\SDNW} class, proposed first by~\cite{maslov2016advantages}. Suppose that we can identify a subcircuit such that the corresponding unitary is of the form $VUV^\dagger$ for a permutation $V$ and $U$ is a controlled unitary such that the qubits on which $V$ is acting are either controls of $U$ or are ignored when $U$ is applied. Consequently, we can replace $V$ with its relative-phase variant $\tilde V$, i.e.~we can replace $VUV^\dagger$ with $\tilde VU\tilde V^\dagger$. This is possible since any phase introduced by $\tilde V$ will later be uncomputed after $\tilde V^\dagger$ is applied:
\begin{equation}
\begin{split}
    \ket{b,c} &\xrightarrow{\tilde V} \exp(i \psi_b) \ket{V(b),c} \xrightarrow{U} \exp(i \psi_b)  \ket{V(b)}U_{V(b)}\ket{c} \\
   & \xrightarrow{\tilde V^\dagger} \exp(i \psi_b) \exp(-i \psi_{V^\dagger(V(b))})  \ket{V^\dagger(V(b))}U_{V(b)}\ket{c}  \\
   & = \exp(i \psi_b) \exp(-i \psi_{b})  \ket{b}U_{V(b)}\ket{c} \\
   & =  \ket{b}U_{V(b)}\ket{c},
\end{split}
\end{equation}
where we used the fact that $V^\dagger(V(b)) \equiv b$ and when applying $\tilde V^\dagger$ on the quantum state, the inverse of the phase is applied. For convenience, we use the same notation for the unitary and the corresponding permutation, so that $V(b)$ denotes the application of permutation $V$ on $b$. In Fig.~\ref{fig:vuvdagger}, subcircuits corresponding to such $VUV^\dagger$ are identified and marked by rectangles.

\begin{figure}[t]
\centering
\includegraphics[scale = .72]{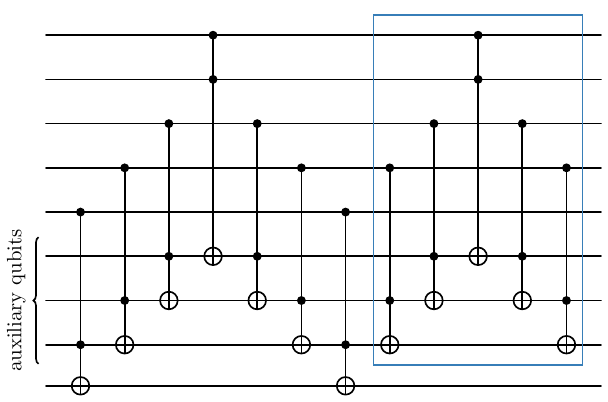}\hfill
\includegraphics[scale = .69]{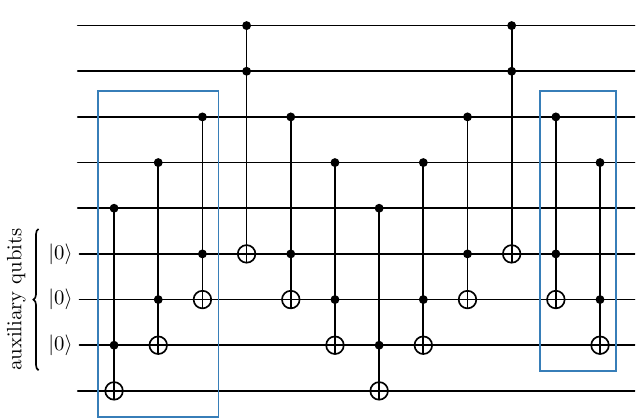}
\caption{(left) Since all the gates inside the rectangle are controls on the main qubits, we can remove them as part of the dirty or clean wasting-entangled preserving transformation. (right)
Given that the auxiliary qubits are initialized to be in state $\ket{0}$, the gates inside the rectangles can be removed, resulting in a {\SCNW} preserving transformation. Note that only the gates from the first rectangle can be removed for the wasting preserving transformation.} 
\label{fig:dirty-clean}
\end{figure}

Let us now introduce a wasting-entangled preserving transformation, in which recovering the state of some of the auxiliary qubits is not required.
For such wasted qubits, we can remove the gates at the end that do not interfere with qubits whose states should be preserved.
Equivalently, we can remove all the controlled unitary operation $U$ for which the qubits from the main system are either not tackled or among the controls of $U$.
After the removal of such gates, we will end up in a clean or dirty wasting-entangled scenario. This is visualized in Fig.~\ref{fig:dirty-clean} (left).
Note that the removal of the gates inside the rectangle results in the dirty wasting-entangled scenario. 

If auxiliary qubits are guaranteed to be in state $\ket{0}$ initially (an assumption for clean classes), we can propose further transformations in the presence of controlled unitaries appearing at the beginning of the circuit controlled by auxiliary qubits. In such a case, the controlled unitary can be replaced by $U_0$, the unitary that acts on the target qubits when the control is in state $\ket{0}$, check Eq.~\eqref{eq:controlled-def}. In an extreme case, if $U_0$ is the identity, one can remove the gate completely. Such a transformation preserves any clean class, as the application of $U_0$ is equivalent to the original controlled unitary. In the clean non-wasting case, we can apply this to controlled unitaries that appear at the end of the circuit as well, as those qubits are guaranteed to end up in state $\ket{0}$ at the end of the circuit. This is visualized in Fig.~\ref{fig:dirty-clean} (right), where the gates inside the rectangles can be safely removed to produce a clean non-wasting implementation. Note that in this example, the removal of the Toffoli gates is based on interpreting them as controlled unitaries with an auxiliary qubit being the single control qubit and the unitary operation being a CNOT. Note that since dirty implementations are special cases of the clean ones, this way they can be easily converted to the clean ones.

\begin{figure}[t]
\centering
\includegraphics[scale = .8]{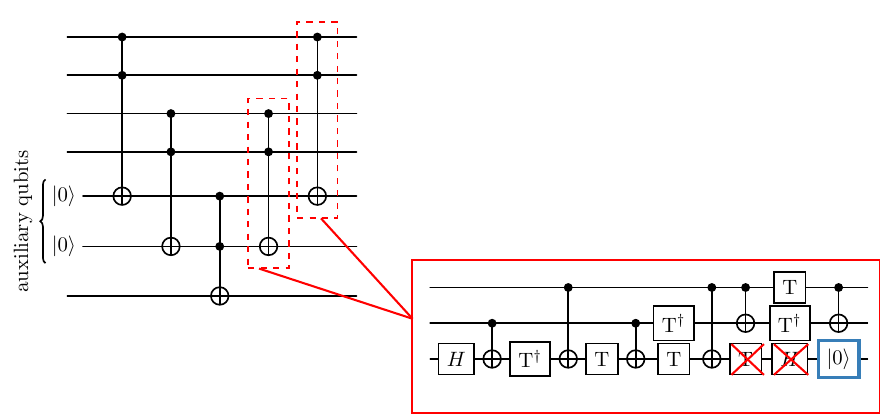}
\caption{The $T$ and $H$ gates appearing at the end for both Toffoli gates can be removed from the circuit, resulting in a clean wasting-separable preserving transformation. The reset gate applied at the end results in a clean non-wasting implementation.} 
\label{fig:reset}
\end{figure}

Next, we consider two transformations, the first being clean wasting-separable classes preserving and the second being clean non-wasting classes preserving. In the first case, one is allowed to remove any gates from the end that act solely on the auxiliary system. In the example provided in Fig.~\ref{fig:reset}, we manage to save two $T$ gates from the circuit (one from each Toffoli inside dashed rectangles), a gate that is considered to be costly for fault-tolerant computing. The second transformation in addition introduces reset gates  -- in this case, after removing the gates as in the previous transformation, in addition, we apply reset operations on the auxiliary qubits at the end of the circuit. Note that this way we can transform clean wasting-separable implementations into clean non-wasting by just adding reset operations at the end of the circuit, as visualized in Fig.~\ref{fig:reset}. 

\begin{figure}[h]

         \includegraphics[height = 5.5cm]{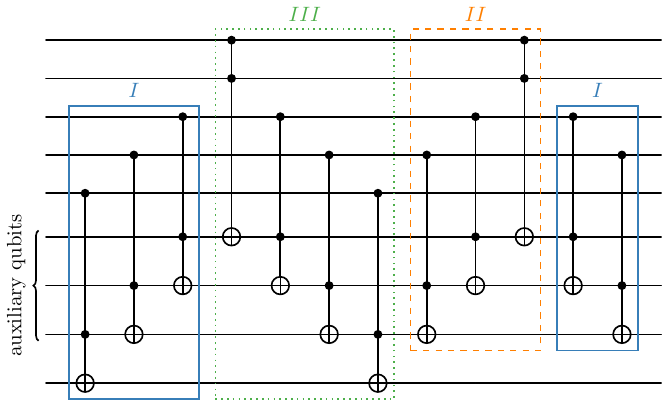}
     \hfill
         \includegraphics[height = 5.1cm]{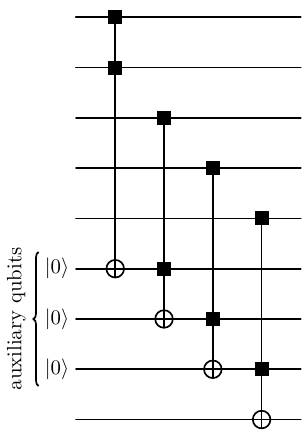}
             
   \caption{The circuit on the left-hand side is a dirty non-wasting implementation of MCT. Following the three steps as explained at the end of Sec.~\ref{sec:transformations}, we end up with a clean wasting-entangled implementation, which is shown on the right-hand side. The three-qubit gate with full-black square symbolizes the relative-phase Toffoli gate from~\cite{PhysRevA.52.3457}. }
   \label{fig:newmct}%

\end{figure}

We end this section by creating a new simple MCT implementation using the transformations above. We will construct a member of the clean wasting-entangled class. The process and its final outcome are shown in Fig.~\ref{fig:newmct}. We start with the dirty non-wasting implementation of MCT~\cite{PhysRevA.52.3457}. First, we remove the left-most three and right-most two Toffoli operations using clean non-wasting preserving transformation, which is marked with the blue, solid line rectangles. We are left with an implementation that is equivalent to the decomposition introduced previously in~\cite{maslov2016advantages}. Since the remaining three right-most Toffolis inside the orange dashed rectangle are controls on the main system, they can be removed resulting in a wasting-entangled implementation. Finally, as we showed before, there is no difference between strict and relative-phase wasting-entangled classes, so we can replace all the remaining Toffolis inside the dotted green rectangle with their relative-phase versions. The resulting circuit is depicted in Fig.~\ref{fig:newmct} (right). 

\begin{figure}[h]
	
	\includegraphics[height = 5.5cm]{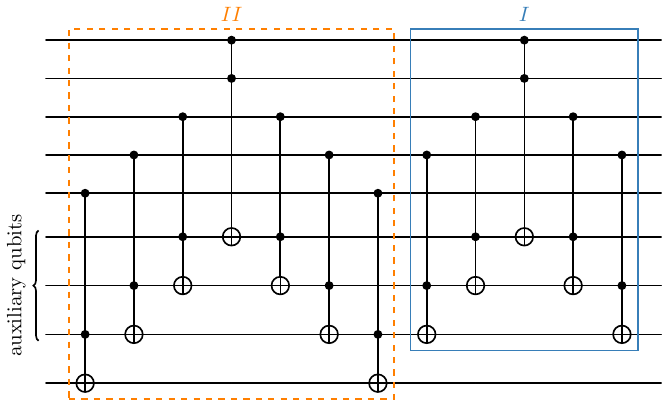}
	\hfill
	\includegraphics[height = 5.1cm]{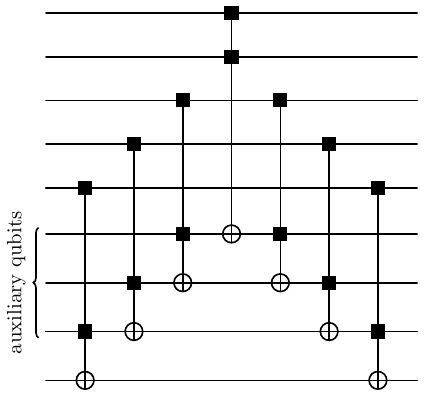}

	\caption{The circuit on the left-hand side is a dirty non-wasting implementation of MCT. Following the two steps as explained at the end of Sec.~\ref{sec:transformations}, we end up with a dirty wasting-entangled implementation, which is shown on the right-hand side. The three-qubit gate with full-black square symbolizes the relative-phase Toffoli gate from~\cite{PhysRevA.52.3457}. }
	\label{fig:newmct_dirty}%
	
\end{figure}

Note that one can construct in a very similar way, a member of the dirty wasting-entangled class. The process and its final outcome are shown in Fig.~\ref{fig:newmct_dirty}. To construct such a decomposition, we start by removing the last 5 Toffoli gates, as it is a controlled unitary with controls on the main qubits. Then, we replace all the remaining Toffoli gates with their relative-phase variants. The resulting circuit is depicted in Fig.~\ref{fig:newmct_dirty} (right). 

\section{Conclusions and Discussion}\label{sec:conclusions}

In this paper, we presented a comprehensive classification of unitary permutations and the quantum circuits that implement them. This classification is based on several factors: whether the input state of the auxiliary system is predefined to be $\ket{0}$, whether the state of the auxiliary system is preserved at the end of the computation, and whether the implementation introduces local phases. Through this classification, we identified 12 different classes, however, we showed that 2 of them were redundant. In addition, to make our findings more rigorous, we provided a formal definition for each class, which can be used to verify algorithmically whether a given quantum circuit belongs to a particular class. Finally, we presented a partial order of the classes introduced through the inclusion of sets containing all the unitaries (or quantum circuits) belonging to the particular class.  

Our classification applies to general permutations, however, since the multi-controlled Toffoli gates are the most commonly used gate in the literature, we made a brief review of their implementations and categorized them based on our classification. We found out that despite the multiple implementations available, there exist implementations for only 3 classes, which is rather limited considering that we identified 10 distinct classes. To address this limitation, we introduced the concept of transformations, which can be seen as a set of rules modifying the quantum circuit in a way that preserves a particular class. This way we were able to simplify existing implementations so that the new ones use fewer resources, like the number of gates or depth. 

While we do not specify how to implement a circuit for a particular permutation, we offer a framework to construct quantum circuits using potentially fewer resources. By repetitively applying the transformations proposed in the paper to particular implementations of clean non-wasting or dirty non-wasting circuits, one can efficiently generate quantum circuits that are members of larger classes. Since larger (in inclusion) classes contain more unitaries and quantum circuits, the chances of obtaining a more efficient implementation than the one initially prepared are higher. This opens up room for new heuristic transpilation techniques, especially for the circuits for which it is possible to identify qubits that can be wasted. For instance, let us consider a scenario in which there are two permutations applied in a row that share some auxiliary qubits. One can choose between wasting for the former permutation and dirty for the latter one, or non-wasting for the former and clean for the latter, depending on which scenario turns out to be more beneficial.

For completeness, we implemented some of the MCT decompositions and the verifications for all the classes introduced in the paper. The software includes also some basic transformations considered in this paper. The relevant source code can be found in \url{https://github.com/QuCoNot/QuCoNot}.

\paragraph{Acknowledgements}

A.G. has been partially supported by National Science Center under grant agreements 2019/33/B/ST6/02011, and 2020/37/N/ST6/02220. \"O.S. acknowledges support from National Science Center under grant agreement 2019/33/B/ST6/02011. The authors would like to thank Zolt\'an Zimbor\'as and Jarosław Miszczak for discussions on the results presented in this paper.

We acknowledge QWorld Association for organizing the remote internship program QIntern 2022 during which we initiated the work presented in this paper.

\bibliographystyle{ieeetr}
\bibliography{mct}

\appendix
\section{Reverse Kronecker product for untaries}\label{sec:reverse-kronecker-product}

In this section, we will make some remarks on the reverse Kronecker product. In particular, we will show that given a unitary $U$, if decomposition $U=V\otimes W$ exists, then one can always find such $V$ and $W$ that are unitary operations. We assume that we decompose unitary $U$ into $V\in L(\C^{n_V})$ and $W\in L(\C^{n_W})$ for arbitrary fixed $n_V>0$ and $n_W>0$. In the rest of the discussion, whenever we refer to the Kronecker product of $A \otimes B$ or $\ket{\varphi}\otimes \ket{\psi}$, it should always be understood within the context of the product space $L(\C^{n_V})\otimes L(\C^{n_W})$ or $\C^{n_V}\otimes \C^{n_W}$, respectively.

Let us first show that the only valid decomposition for identity is $I = aI \otimes \frac{1}{a}I$ for $a\neq 0$. Suppose we have an arbitrary decomposition $I=A\otimes B$. First of all, by the very definition of the Kronecker product, both $A$ and $B$ are diagonal. Let $\ket{b}\in \C^{n_W}$ be an arbitrary standard basis vector. Then for any basis vector $\ket{b'}\in \C^{n_V}$  
\begin{equation}
 1 = \bra{b',b}I\ket{b',b} = \bra{b'}A \ket{b'} \bra{b}B \ket{b}  = \bra{b'} (\bra{b}B \ket{b} A) \ket{b'}.
\end{equation}
Because of this, and also since $A$ is diagonal, $A' \coloneqq \bra{b}B \ket{b}  A$ is the identity matrix. Therefore $A=A'/\bra{b}B \ket{b}$ must be of the form $aI$ for some $a\neq 0$. Similarly one can show that $B=a'I$. Then it follows
\begin{equation}
    I = (aI) \otimes (a'I) = (aa') I
\end{equation}
thus $a' = \frac{1}{a}$.

Let us show that if $U=A_1\otimes A_2$ exists, then both $A_1$ and $A_2$ are rescaled unitaries. Indeed one can see that
\begin{equation}
I = U U^\dagger = (A_1 \otimes A_2) (A_1^\dagger \otimes A_2^\dagger) = (A_1A_1^\dagger \otimes A_2A_2^\dagger).
\end{equation}
In light of previous reasoning, we have that $A_1A_1^\dagger=aI$ and $A_2A_2^\dagger=\frac{1}{a}I$ for some $a\neq 0$. Thus $A_1' = \frac{1}{\sqrt a} A_1$ and $A_2' = \sqrt a A_2$ are unitaries, and therefore $A_1,A_2$ are indeed rescaled unitaries.

Finally, let $U=V'\otimes W'$ be any decomposition with $V',W'$ being rescaled unitaries, i.e. $V'= \frac{1}{\sqrt{a}}V$ and $W'= \sqrt{a}W$ for some unitary $V,W$ and $a\neq0$. Note 
\begin{equation}
    V\otimes W = \sqrt a V' \otimes \frac{1}{\sqrt a}W' = V' \otimes W' =U,
\end{equation}
which makes $U=V\otimes W$ a valid decomposition of $U$ into Kronecker product of unitaries.

\paragraph{How to find a decomposition into a tensor product of unitaries}
Using results from~\cite{van1993approximation}, one can show that after rearranging the elements of $U$, the problem can be reduced to finding the vector corresponding to the largest singular value. Thus (given polynomial size of $U$) the decomposition into $V'$ and $W'$ can be found efficiently. As a particular rescaling parameter, one can take $a=(\det W')^2$, since
\begin{equation}
    \det(W') = \sqrt a \det W = \sqrt a \exp(i \alpha)
\end{equation}
for some real $\alpha$. This procedure is implemented in \url{https://github.com/QuCoNot/QuCoNot} in the file \emph{\text{reverse\_kronecker\_product.py}}.

\end{document}